\documentclass[aps,prc,twocolumn,showpacs,superscriptaddress]{revtex4-1}
\usepackage{graphicx}% Include figure files
\usepackage{dcolumn}% Align table columns on decimal point
\usepackage{bm}% bold math
\usepackage{epstopdf}
\usepackage{multirow}
\usepackage{enumitem}
\usepackage{paralist}
\usepackage{booktabs}
\usepackage[version=3]{mhchem}
\usepackage{color}
\usepackage{placeins}
\usepackage{verbatim}
\begin{document}

\preprint{APS/123-QED}

\title{High-precision quadrupole moment reveals significant intruder component in $^{33}_{13}$Al$_{20}$  ground state}

\author{H. Heylen}
\affiliation{KU Leuven, Instituut voor Kern- en Stralingsfysica, 3001 Leuven, Belgium}
\author{M. De Rydt}
\affiliation{KU Leuven, Instituut voor Kern- en Stralingsfysica, 3001 Leuven, Belgium}
\author{G. Neyens}
\email{gerda.neyens@kuleuven.be}
\affiliation{KU Leuven, Instituut voor Kern- en Stralingsfysica, 3001 Leuven, Belgium}
\author{M.L.~Bissell}
\affiliation{KU Leuven, Instituut voor Kern- en Stralingsfysica, 3001 Leuven, Belgium}
\author{L.~Caceres}
\affiliation{Grand Acc\'el\'erateur National d'Ions Lourds (GANIL), CEA/DRF-CNRS/IN2P3, B.P. 55027, F-14076 Caen Cedex 5, France}
\author{R.~Chevrier}
\affiliation{CEA, DAM, DIF, F-91297 Arpajon Cedex, France}
\author{J.M.~Daugas}
\affiliation{CEA, DAM, DIF, F-91297 Arpajon Cedex, France}
\author{Y.~Ichikawa}
\altaffiliation[Present address: ]{Department of Physics, Tokyo Institute of Technology, 2-12-1 Oh-okayama, Meguro-ku, Tokyo 152-8551, Japan}
\affiliation{RIKEN Nishina Center, 2-1 Hirosawa, Wako, Saitama 351-0198, Japan}
\author{Y.~Ishibashi}
\affiliation{RIKEN Nishina Center, 2-1 Hirosawa, Wako, Saitama 351-0198, Japan}
\affiliation{Institute of Physics, University of Tsukuba, 1-1-1 Tennodai, Tsukuba, Ibaraki 305-8577, Japan}
\author{O.~Kamalou}
\affiliation{Grand Acc\'el\'erateur National d'Ions Lourds (GANIL), CEA/DRF-CNRS/IN2P3, B.P. 55027, F-14076 Caen Cedex 5, France}
\author{T.J.~Mertzimekis}
\affiliation{Department of Physics, University of Athens, Zografou Campus, GR-15784, Athens, Greece}
\author{P. Morel}
\affiliation{CEA, DAM, DIF, F-91297 Arpajon Cedex, France}
\author{J.~Papuga}
\affiliation{KU Leuven, Instituut voor Kern- en Stralingsfysica, 3001 Leuven, Belgium}
\author{A.~Poves}
\affiliation{Departamento de F\'isica Te\'orica and IFT-UAM/CSIC, Universidad Aut\'onoma de Madrid, E-28049 Madrid, Spain}
\affiliation{KU Leuven, Instituut voor Kern- en Stralingsfysica, 3001 Leuven, Belgium}
\author{M.M.~Rajabali}
\affiliation{KU Leuven, Instituut voor Kern- en Stralingsfysica, 3001 Leuven, Belgium}
\author{C. St\"odel}
\affiliation{Grand Acc\'el\'erateur National d'Ions Lourds (GANIL), CEA/DRF-CNRS/IN2P3, B.P. 55027, F-14076 Caen Cedex 5, France}
\author{J.C.~Thomas}
\affiliation{Grand Acc\'el\'erateur National d'Ions Lourds (GANIL), CEA/DRF-CNRS/IN2P3, B.P. 55027, F-14076 Caen Cedex 5, France}
\author{H.~Ueno}
\affiliation{RIKEN Nishina Center, 2-1 Hirosawa, Wako, Saitama 351-0198, Japan}
\author{Y.~Utsuno}
\affiliation{Advanced Science Research Center, Japan Atomic Energy Agency, Tokai, Ibaraki 319-1195, Japan}
\affiliation{Center for Nuclear Study, University of Tokyo, Hongo, Bunkyo-ku, Tokyo 113-0033, Japan}
\author{N.~Yoshida}
\affiliation{Department of Physics, Tokyo Institute of Technology, 2-12-1 Oh-okayama, Meguro-ku, Tokyo 152-8551, Japan}
\author{A.~Yoshimi}
\altaffiliation[Present address: ]{Research Core for Extreme Quantum World, Okayama University, Okayama 700-8530, Japan}
\affiliation{RIKEN Nishina Center, 2-1 Hirosawa, Wako, Saitama 351-0198, Japan}

\date{\today}

\begin{abstract}
The electric quadrupole moment of the $^{33}_{13}$Al$^{}_{20}$ ground state, located at the border of the island of inversion, was obtained using continuous-beam $\beta$-detected nuclear quadrupole resonance ($\beta$-NQR). From the measured quadrupole coupling constant \mbox{$\nu_Q = 2.31(4)$ MHz} in an \mbox{$\alpha$-Al$_2$O$_3$} crystal, a precise value for the electric quadrupole moment is extracted: $\left|Q_s\left(^{33}\text{Al}\right)\right|=141(3)$ mb. A comparison with large-scale shell model calculations shows that $^{33}$Al has at least 50\% intruder configurations in the ground state wave function, favoring the excitation of two neutrons across the $N=20$ shell gap. $^{33}$Al therefore clearly marks the gradual transition north of the deformed Na and Mg nuclei towards the normal $Z\geq14$ isotopes.

\end{abstract}

%Change
\maketitle

%%% INTRODUCTION
\section{Introduction}
In the last decades, the study of isotopes far from \mbox{$\beta$-stability} has decisively altered our understanding of the nuclear shell structure through the observation of many phenomena at variance with conventional expectations. One of the oldest examples is the region of deformation around the classic magic number $N=20$, which was first discovered through anomalies in the properties of neutron-rich Na and Mg isotopes \cite{Thibault1975, Huber1978, Detraz1979}. This deformation is associated with two particle-two hole (2p-2h) neutron excitations across $N=20$, from the $sd$ orbitals to the $fp$ orbitals. Due to the combination of a reduced $N=20$ shell gap and a large gain in quadrupole correlation energy, the energy for such particle-hole configurations is lowered drastically and, at specific $N,Z$-combinations, these so-called intruder configurations can even become the ground state, below the normal $sd$ configurations. The isotopes for which this inversion of normal and intruder configurations occurs are referred to as belonging to the island of inversion. 
\\ Warburton et al. \cite{Warburton1990} originally predicted that the island of inversion around $N=20$ contains the $_{10}$Ne, $_{11}$Na and $_{12}$Mg isotopes with \mbox{$20 \leq N \leq 22$}. Since then, an extensive research program has focused on the characterization of the low-energy properties of these isotopes and their neighbors, in order to understand the origin of the changing nuclear structure \cite{Utsuno2004,Neyens2005,Yordanov2007,Himpe2008,Yordanov2012,Chaudhuri2013}. Moreover, it is recognized  that the development of deformation happens gradually as a function of $N$ and $Z$ \cite{Neyens2007}. Since nuclei at the edge of the island of inversion have a transitional character with a varying amount of normal and intruder configurations, their study is imperative to understand the drivers of the sudden structural changes in the region.
\\ The neutron-rich Al isotopes ($Z=13$) constitute the northern border of the island of inversion, between the spherical Si isotopes and the deformed Mg isotopes. $^{33}_{13}$Al$_{20}$ is thought to be a key isotope as the transition into the island is particularly rapid in the $N=20$ isotones. In $^{34}_{14}$Si$_{20}$ a 2p-2h intruder state was recently found at 2719~keV \cite{Rotaru2012} which comes down almost 4~MeV in $^{32}_{12}$Mg$_{20}$ where it becomes the ground state lying 1058~keV below the normal configuration \cite{Wimmer2010}. $^{33}$Al is therefore expected to have a transitional nature, with potentially a significant amount of intruder admixtures in its ground state. Experimental measurements to determine this intruder component have led to conflicting results. Whereas mass measurements \cite{Kwiatkowski2015} and the $^{33}$Al $\rightarrow{}^{33}$Si $\beta$-decay study \cite{Morton2002} place $^{33}$Al outside the island of inversion, the \mbox{$g$-factor \cite{Himpe2006}} and the (debated) $^{33}$Mg  $\rightarrow{}^{33}$Al $\beta$-decay scheme \cite{Tripathi2008,Yordanov2010,Tripathi2010} suggest a sizable intruder component in the $^{33}$Al ground state, while also one-neutron removal cross sections and longitudinal momentum distributions do not exclude the presence of intruder configurations \cite{Nociforo2012}. The quadrupole moment is a particularly well-suited observable to shed light on this issue due to its sensitivity to E2 collectivity and 2p-2h excitations in the wave function. However, as the precision on the previous measurement was low (12\%), no firm conclusion could be drawn \cite{Shimada2012}.
\\ Here, we present the first high-precision quadrupole moment measurement of the $^{33}$Al ground state ($I = 5/2^+$, $t_{1/2}$ = 41.7 ms) via continuous-beam $\beta$-NQR at \mbox{LISE-GANIL}. Via a comparison with large-scale shell model calculations, the mixing of intruder configurations in the ground state wave function of $^{33}$Al is investigated.

\section{Experimental details}
%%% EXPERIMENTAL RESULTS
The $\beta$-detected nuclear quadrupole resonance \mbox{($\beta$-NQR)} measurements were performed at the LISE-GANIL facility. Neutron-rich $^{33}$Al nuclei were produced and spin-polarized in a projectile fragmentation reaction induced by a $^{36}$S$^{16+}$ primary beam (77~MeV/u and 1.8~$\mu$A) on a 1018 $\mu$m thick $^{9}$Be target. After selection in the LISE3 fragment separator, a 94\% pure beam of $^{33}$Al ($8 \cdot 10^3$ particles per second) was guided to the $\beta$-NMR/NQR set-up \cite{Derydt2009b}. Here the beam was implanted into a $\alpha$-Al$_2$O$_3$ single crystal at room temperature. The crystal was placed in a static external magnetic field ($B_0$ = 2002(7) gauss) along the spin direction in order to maintain the spin-polarization and induce a Zeeman splitting. Perpendicular to the static field, a radio frequency field with frequency $\nu_\text{rf}$ was applied, generated by the amplified rf generator signals sent to two Helmholtz coils mounted around the crystal. The $\beta$-decay of the radioactive $^{33}$Al isotopes was detected in two sets of $\Delta E-E$ plastic scintillator telescopes placed above and below the implantation crystal. To reduce the influence of scattering and noise events, $\Delta E-E$ scintillator coincidences were required. For a spin-polarized ensemble, the $\beta$-decay is anisotropic in space and the asymmetry $\mathcal{A}$ is defined as 
\begin{equation*}
\small
\mathcal{A} = \frac{N_\text{up} - N_\text{down}}{N_\text{up} + N_\text{down}} \approx A_\beta P
\normalsize
\end{equation*}
with $A_\beta$ a parameter determined by the $\beta$-decay properties and $P$ the amount of spin-polarization. $N_\text{up}$ and $N_\text{down}$ are the number of counts detected in the upper and lower detectors, respectively. 
\\ The implanted $^{33}$Al nuclei are exposed to an axial symmetric electric field gradient $V_{zz}$ in the hexagonal \mbox{$\alpha$-Al$_2$O$_3$}. In addition to the Zeeman splitting due to the external magnetic field, this field gradient induces a shift due to the quadrupole interaction. In a field of $\sim$2000 gauss, the magnetic interaction dominates the electric quadrupole interaction and the splitting of the $|I, m \rangle$-states is in first-order given by \cite{Abragam1961}
\begin{align*}
\small
\nu_m - \nu_{m+1} = \nu_L - \frac{3\nu_Q}{8I(2I-1)}(2m+1)(3 \cos^2 \theta -1) 
\normalsize
\end{align*}
Here $\theta$ is the angle between the external field and the symmetry axis of the electric field gradient in the crystal, which was put to zero in the experiment. 
From a precise measurement of the transition frequencies, the nuclear $g$-factor and spectroscopic quadrupole moment $Q_s$ can be extraced via the relations  for the Larmor frequency $\nu_L = |g|\mu_NB_0/h$ and the quadrupole coupling constant $\nu_Q=e|Q_s|V_{zz}/h$.
	\begin{figure}[t]
	\includegraphics[width=0.9\columnwidth]{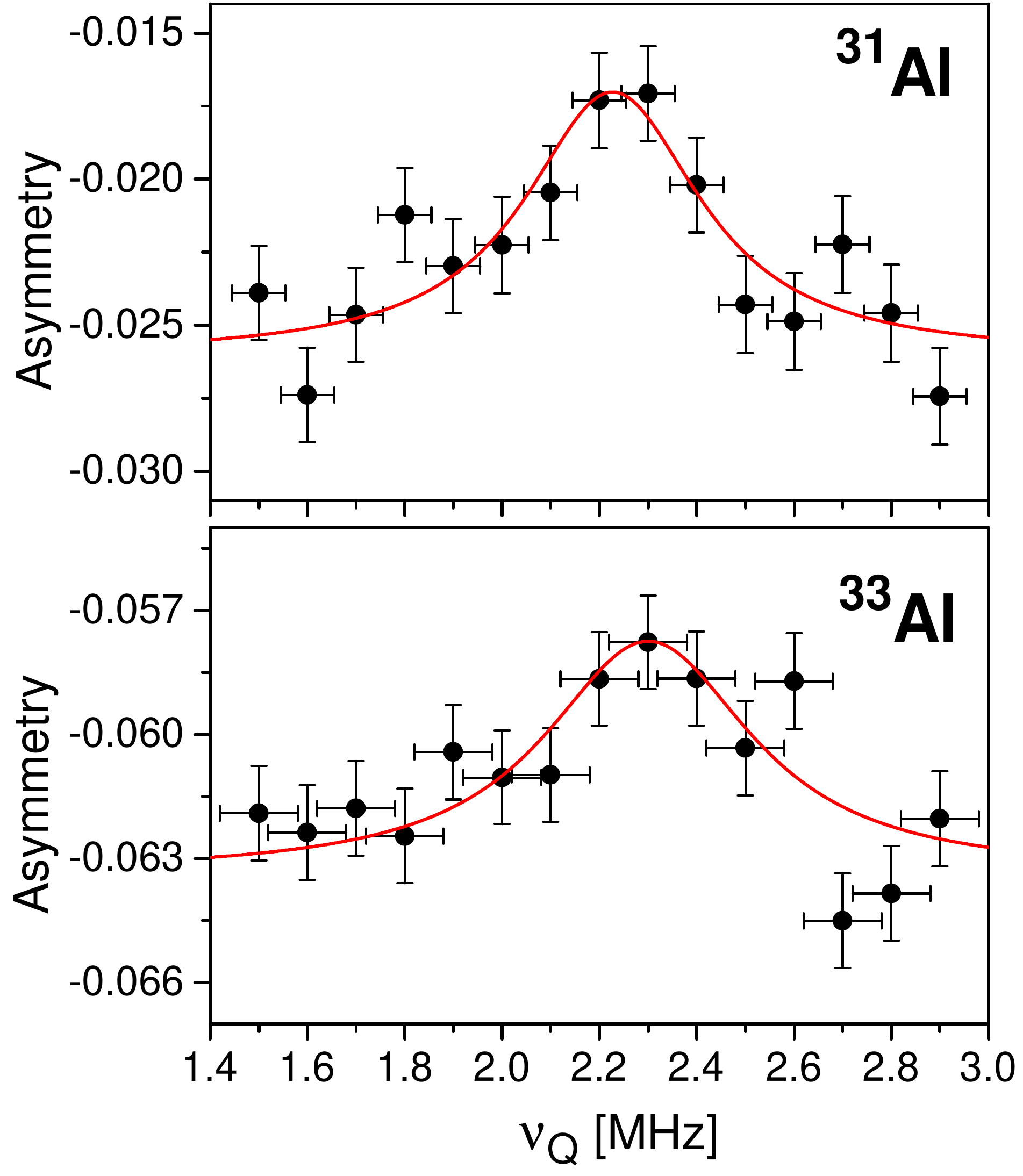}
	\caption{Experimental $\beta$-NQR spectra for the $^{31}$Al and $^{33}$Al ground states with a frequency modulation of respectively 55~kHz and 80~kHz.} 
	\label{Fig:spectra_33Al}
	\end{figure}
\\ For a $I = 5/2^+$ nucleus such as $^{33}$Al, there exist five transition frequencies between the $|I, m \rangle$-states \mbox{($m = -5/2, -3/2, \ldots,\  5/2$)}. When the applied rf-field has a frequency that matches one of these frequencies, the population of the two magnetic substates is equalized, which induces only a partial destruction of the ensemble polarization, reflected as a small change in the asymmetry $\mathcal{A}$ \cite{Borremans2005}. In order to enhance this change in asymmetry and thus the resonance signal, the polarization needs to be fully destroyed. As the five transition frequencies are correlated through the common coupling constant $\nu_Q$, they can be scanned simultaneously by varying this $\nu_Q$.  When the resonance condition for $\nu_Q$ is met, the population of all $|I, m \rangle$-states is equalized, and a full change in asymmetry is observed. To guarantee all five frequencies are at resonance simultaneously, precise knowledge of $\nu_L$ is required. Therefore,  prior to the $\beta$-NQR measurement, a $\beta$-NMR measurement in a cubic Si crystal was performed yielding $\nu_L$ in the applied $B_0$. During the $\beta$-NMR and NQR measurements, the rf-frequencies were scanned in discrete steps and for each frequency, a frequency modulation of more than 50\% of the step size was applied to cover the full frequency range. A detailed description of the methodology of continuous-beam $\beta$-NQR can be found in \cite{Derydt2009b}. 
\\ The set-up was first tested using the well-known $^{31}$Al isotope after which the $^{33}$Al measurements were performed in similar conditions. Typical spectra are shown in Fig.~\ref{Fig:spectra_33Al} and an overview of the results can be found in Table~\ref{Table:results}. The quadrupole moments of $^{31}$Al and $^{33}$Al have been extracted from the measured quadrupole coupling constant, relative to that of $^{27}$Al
\begin{equation*}
\small
|Q_s(^A\text{Al})| = \frac{ \left|Q_s(^{27}\text{Al}) \right| \nu_Q(^{A}\text{Al})}{\nu_Q(^{27}\text{Al})}
\normalsize
 \end{equation*}
 The $^{31,33}$Al results are in good agreement with the previously known literature values, and the precision on $\left|Q_s\left(^{33}\text{Al}\right)\right|$ is improved with more than a factor of 5,  from 12\% in the previous measurement \cite{Shimada2012} to 2\% in this work. This precision allows a detailed discussion of the intruder components in the $^{33}$Al ground state.
\begin{table}[t!]  % for columnwide tables
\caption{The quadrupole coupling constant of the $^{31,33}$Al ground state extracted during the $\beta$-NQR measurements in \mbox{$\alpha$-Al$_2$O$_3$}. The results are compared to the literature values and for completeness also the $^{27}$Al reference values are presented.}
\label{Table:results}
\begin{ruledtabular}
\begin{tabular}{l l l l l l l l l }
\\ [-2.3ex]
 & $N$ & $I^\pi$ & $t_{1/2}$ & $\nu_Q$ [MHz] & $Q_s$ [mb] & Ref. \\
 [0.45ex]
\colrule \\ [-1.85ex]
$^{27}$Al & 14 & 5/2$^+$ & stable &  2.4031(2) & +146.6(10) & \cite{Derydt2013,Kello1999} \\
$^{31}$Al & 18 & 5/2$^+$ & 644 ms & 2.24(3)  & 136.5(23) & This work \\
 & & & &2.19(2) & 134.0(16) &  \cite{Derydt2009a}\\
 $^{33}$Al & 20 & 5/2$^+$ & 41.7 ms &  2.31(4) & 141(3) & This work \\
  & & & & 2.2(3) &  132(16) & \cite{Shimada2012}\\
\end{tabular}
\end{ruledtabular}
\end{table}

\section{Results and discussion}
\begin{figure}[t]
	\includegraphics[width=\columnwidth]{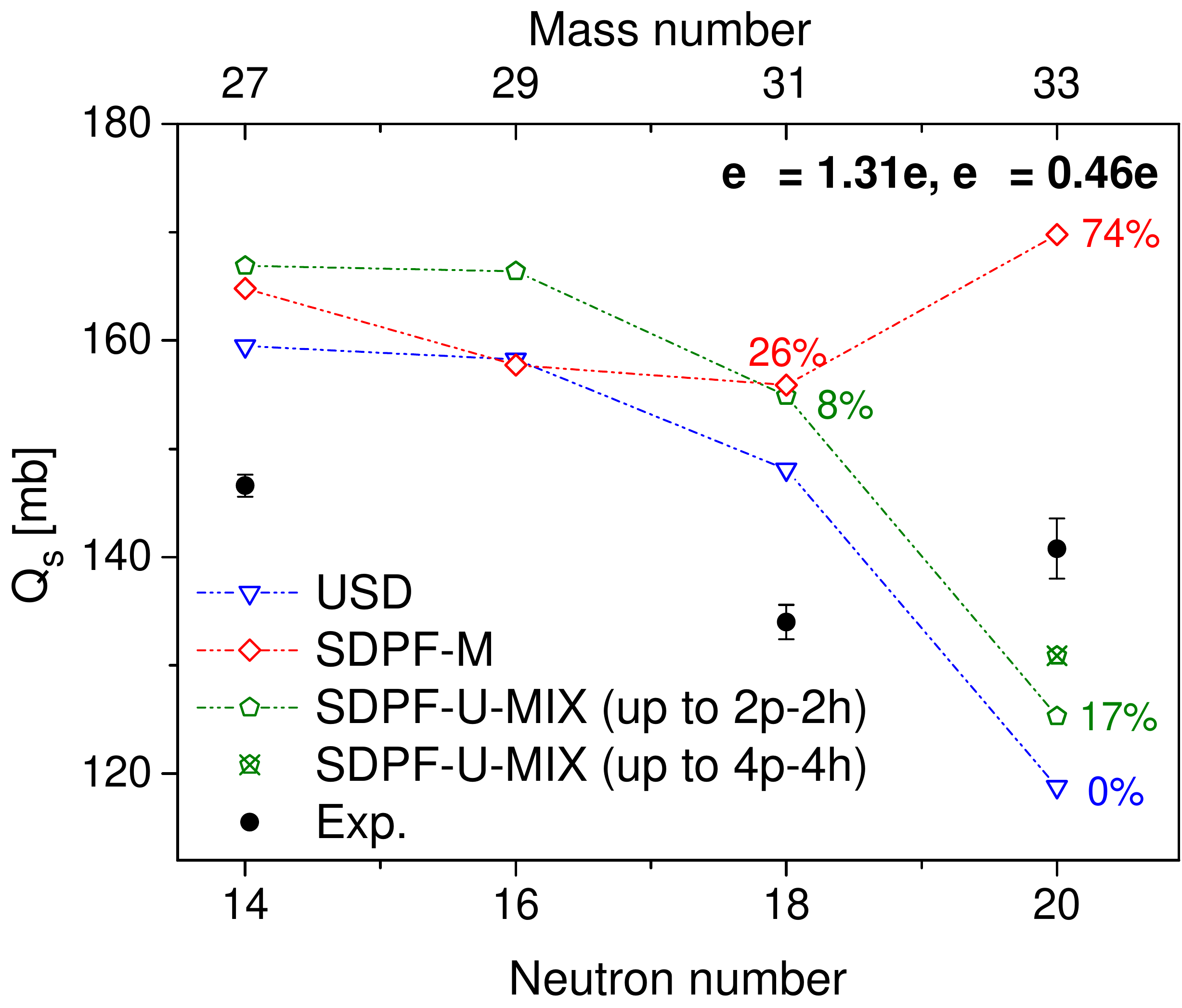}
	\caption{The experimental quadrupole moments compared to shell model calculations using various interactions and the universal effective charges \cite{DufourZuker1996}. The predicted amount of intruder configurations in the wave functions are indicated in brackets.} 
	\label{Fig:Q-int-eff}
	\end{figure}

The experimental ground state quadrupole moment of $^{33}$Al$_{20}$ is very similar to that of $^{27}$Al$_{14}$ and $^{31}$Al$_{18}$, as shown in Fig.~\ref{Fig:Q-int-eff}. This is inconsistent with a picture in which no particle-hole excitations occur: for a closed-shell isotope like $^{33}$Al$_{20}$ the quadrupole moment should decrease with respect to the open-neutron shell isotopes $^{27,31}$Al. That is illustrated in the shell model calculations using the USD shell model interaction \cite{BrownWildenthal1988} with a $^{16}$O core and protons and neutrons restricted to the $sd$-shell model space, the triangles in Fig.~\ref{Fig:Q-int-eff}. Indeed, the $^{33}$Al quadrupole moment would then be entirely due to the proton configuration because the $N=20$ shell closure inhibits all neutron contributions. The open neutron shell in $^{27}$Al and $^{31}$Al on the other hand allows for additional proton-neutron correlations, resulting in an increase in quadrupole moment. The fact the $^{27,31}$Al and $^{33}$Al quadrupole moments are similar in size therefore suggests that this pure $sd$ picture is too limited and that neutron excitations across $N=20$ are necessary to reproduce $Q_s(^{33}$Al).
\\ Therefore, large scale shell model calculations have been performed in different model spaces, using different effective interactions. The results are presented in Fig.~\ref{Fig:Q-int-eff}. The {\sc{Antoine}} shell model code has been used with \mbox{SDPF-U-MIX} \cite{Caurier2014}, the most recent effective interaction for this region, which builds upon a $^{16}$O core and comprises the full $sd$ and $pf$ shell for neutrons. This allows excitations of neutrons across $N=20$ and mixing between $n$p-$n$h configurations. The calculations are truncated to maximum 2p-2h neutron excitations across $N=20$ for Al isotopes up to $^{31}$Al, and up to 4p-4h for $^{33}$Al. As an alternative, the SDPF-M effective interaction \cite{Utsuno1999} is used including excitations from the neutron $sd$-shell to the two lowest levels of the $pf$-shell ($f_{7/2}p_{3/2}$). Previously, calculations with this interaction were performed in the Monte Carlo Shell Model framework \cite{Shimada2012}, while recently full Lanczos diagonalizations (calculated using the {\sc{Mshell64}} code \cite{Mizusaki2016}) have become available, truncated to maximally 4p-4h excitations. Since both diagonalization approaches yield similar results, only the results from the full diagonalization are shown here.
\\ In Fig.~\ref{Fig:Q-int-eff}, all quadrupole moments have been calculated using the theoretically derived universal effective charges $e_\pi = 1.31e$ and $e_\nu = 0.46e$ \cite{DufourZuker1996}. 
The calculations consistently overestimate the experimental quadrupole moments for all three interactions, which is related to the choice of effective charges, as discussed later. At this point it is however more important to note that only the SDPF-M interaction is able to reproduce the experimental trend which goes up at $N=20$. The USD and SDPF-U-MIX interaction on the other hand predict a sizable decrease. Allowing also \mbox{4p-4h} excitations for $^{33}$Al in the \mbox{SDPF-U-MIX} interaction does not really improve the agreement with experiment. An inspection of the intruder admixtures in the $^{33}$Al ground state (indicated in Fig.~\ref{Fig:Q-int-eff}) reveals that 74\% of intruders are predicted using SDPF-M while only 17\% are predicted using \mbox{\mbox{SDPF-U-MIX}}. The discrepancy between the {\mbox{SDPF-U-MIX} calculations and the experimental value at $N=20$ is therefore suggested to at least partly arise from an underestimation of intruders configurations. To obtain a theoretical $^{33}$Al quadrupole moment in better agreement with the experimental trend, the intruder component would need to be doubled, requiring a reduction of the $N=20$ shell gap by 0.5 MeV. However, since the (unmodified) \mbox{SDPF-U-MIX} calculations have been shown to give good results for $E(2^+)$'s, $B(E2)$'s and binding energies in the neighboring isotopic chains \cite{Caurier2014,Kwiatkowski2015}, simply decreasing the $N=20$ shell gap will have detrimental effects on the predictions for these neighboring isotopes and a more careful analysis of the problem is necessary. The observed discrepancy at $^{33}$Al emphasizes the need for high-precision quadrupole moments as a  complementary test for state-of-the-art interactions.
	\begin{figure}[t]
	\includegraphics[width=\columnwidth]{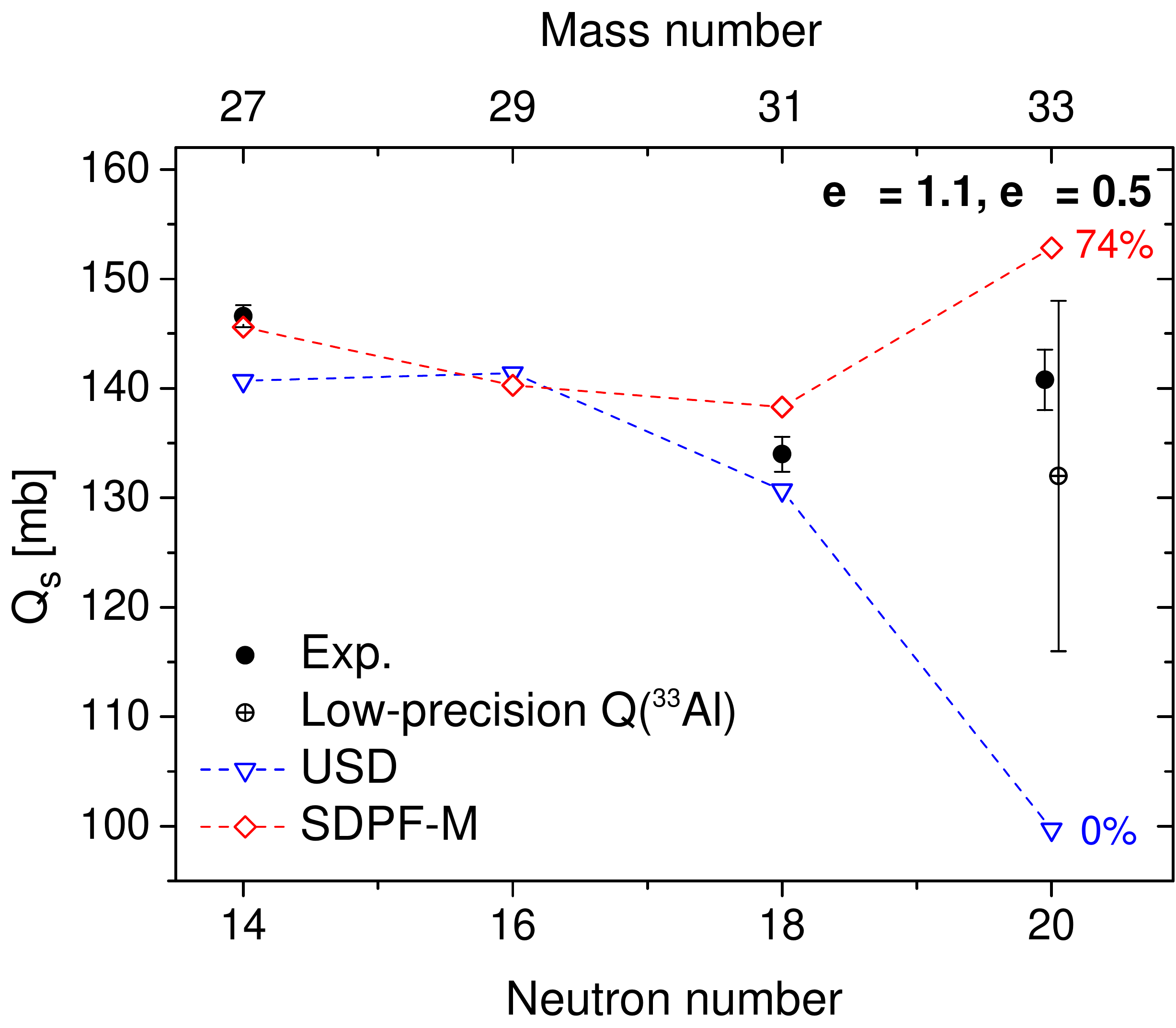}
	\caption{Experimental quadrupole moments of $^{27}$Al, $^{31}$Al and $^{33}$Al where for $^{33}$Al both the high-precision value obtained in this work and the low-precision value \cite{Shimada2012} are shown. These are compared to shell model calculations with the USD and SDPF-M effective interactions using constant effective charges $e_\pi = 1.1e,\ e_\nu = 0.5e$.} 
	\label{Fig:Q-Marieke}
	\end{figure}
\\ To further illustrate the significance of the high-precision result we have obtained in this work, our value is presented in Fig.~\ref{Fig:Q-Marieke} along with the old, low-precision value \cite{Shimada2012} and the results from the USD and the \mbox{SDPF-M} shell model calculations using $e_\pi = 1.1e,\ e_\nu = 0.5e$. The use of these effective charges is justified based on the analysis of the quadrupole moments and $B(E2)$-values of the odd-$Z$ even-$N$ $\pi sd$ isotopes by De Rydt \textit{et al.} \cite{Derydt2009a,PhDDerydt2010} which showed that a reduction of the proton effective charge gives a better reproduction of the experimental data. 
\\ Both USD and \mbox{SDPF-M} can reproduce the experimental quadrupole moments of $^{27,31}$Al, suggesting a structure dominated by normal configurations. For $^{33}$Al, the old low-precision quadrupole moment agrees best with the \mbox{SDPF-M} calculations, even though its large uncertainty prevents firmly ruling out the USD prediction. Only due the high precision obtained in this work, it can be unambiguously concluded that the quadrupole moment of the $^{33}$Al ground state is inconsistent with the USD calculation without intruder configurations. The \mbox{SDPF-M} value on the other hand compares much more favorable to the experimental value, although it is somewhat too high. This suggests that the predicted 74\% intruder configurations is  somewhat overestimated. 
	\begin{figure}[t]
	\includegraphics[width=\columnwidth]{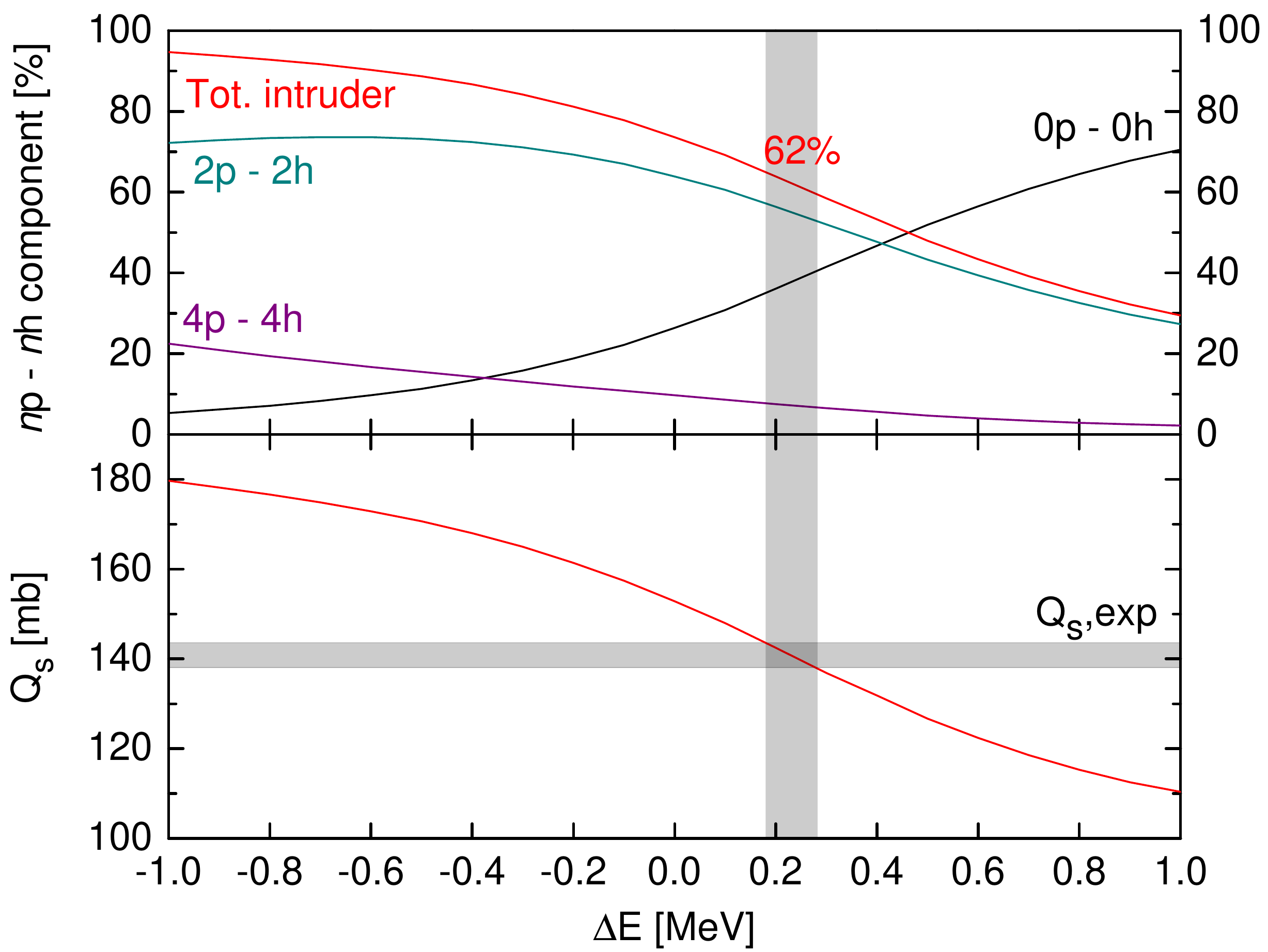}
	\caption{Variation of 0p-0h, 2p-2h and 4p-4h component in the SDPF-M wave function (top) and quadrupole moments (bottom) when the $N=20$ shell gap is changed by $\Delta E$. Quadrupole moments were calculated using the $e_\pi = 1.1e,\ e_\nu = 0.5e$ effective charges. The gray bands indicate the interval around the experimental value.} 
	\label{Fig:Q-influenceN20}
	\end{figure}
\\ The influence of the exact amount of intruder components on the quadrupole moment is therefore investigated through a variation of the size of the $N = 20$ shell gap, obtained by shifting the $\nu f_{7/2}$ and $\nu p_{3/2}$ orbitals simultaneously by $\Delta E$. The results are shown in Fig.~\ref{Fig:Q-influenceN20}. A good agreement between the experimental and theoretical quadrupole moment is already obtained at 62\% intruder configurations, subdivided in 55\% 2p-2h and 7\% 4p-4h components. Immediately, also the clear correlation between the total intruder component in the wave function (in the top panel) and the predicted quadrupole moment (in the bottom panel) becomes evident. Furthermore, the strong dependence of the amount of intruder admixtures on the size of the $N=20$ shell gap is illustrated. Indeed, increasing the shell gap by 1~MeV already reduces the amount intruder configurations by 40\%. 
\\ While the above analysis is instructive to study the sensitivity of the quadrupole moment to the intruder composition of the wave function, it is not meant as a quantitative determination of the optimal $N=20$ shell gap in the SDPF-M interaction which would require a much more elaborate investigation.
\\ Note that the exact amount of intruder admixtures also depends on the choice of effective charges, for example, with the isospin dependent effective charges used previously in \cite{Nagae2009,Shimada2012} 76\% of intruder admixtures would be required to reproduce the experimental quadrupole moment. Whichever effective charges are the most appropriate, it is clear that the main conclusion of this work is solid: to reproduce the quadrupole moment of the $^{33}$Al ground state, a significant admixture of intruder configurations is necessary. This confirms and strengthens the conclusion of the $^{33}$Al $g$-factor measurement \cite{Himpe2006}. However, for a consistent interpretation of the low-energy structure of $^{33}$Al, it would be interesting to investigate how the ground state moments can be consolidated with the apparent contradictory experimental information from the mass and $\beta$-decay measurements \cite{Morton2002,Kwiatkowski2015}.

\section{Conclusion}
%%% CONCLUSION
In conclusion, the ground state quadrupole moment of $^{33}$Al has been measured using the $\beta$-NQR technique at LISE-GANIL. The high-precision experimental value, $|Q_s(^{33}$Al$)|=141(3)$ mb, is inconsistent with a picture in which mixing with neutron intruder configurations beyond $N=20$ is excluded. Furthermore, a comparison with large scale shell model calculations has demonstrated the sensitivity of the $^{33}$Al quadrupole moment to intruder components.  The \mbox{SDPF-U-MIX} interaction predicts only 17\% mixing with intruder configurations but cannot capture the observed rise in quadrupole moment. On the other hand, good agreement between the \mbox{SDPF-M} calculations and the experimental $^{27,31,33}$Al quadrupole moments is obtained, and for $^{33}$Al a ground state configuration mixed with more than 50\% intruders is suggested. Hence, $^{33}$Al cannot be placed outside of the island of inversion but should be considered as an important transitional isotope.

\section*{Acknowledgements}

%%% AKNOWLEDGEMENTS
The authors thank the GANIL staff for their support during the preparation and running of the experiment. This work was partly supported by the European Community FP6 - Structuring the ERA - Integrated Infrastructure Initiative contract EURONS No. RII3-CT-2004-506065, by the FWO-Vlaanderen, by the IAP-programme of the Belgium Science Policy under grand number P6/23 and P7/12, by a grant of the MICINN (Spain) (FPA2011-29854), by the Nupnet network SARFEN (PRI-PIMMNUP-2011-1361), by MINECO (Spain) Centro de Excelencia Severo Ochoa Programme under grant SEV-2012-0249 and by JSPS KAKENHI (Japan) Grant Numbers 21740204 and 15K05094. The experiment was carried out under Experimental Program E437b.

%\bibliography{../../../Al,../../../Mn}% Produces the bibliography via BibTeX.
\bibliography{Al,Mn}% Produces the bibliography via BibTeX.

\end{document}